\documentclass[12pt]{article}

\usepackage{graphicx,a4}

\begin{document}

\centerline{\Large {\bf Periodic Instantons and Quantum--Classical Transitions}}

\centerline{\Large {\bf in Spin Systems}}
\vspace{5mm}
\centerline{J.--Q. Liang\raisebox{0.6ex}{\small *}, 
H.J.W. M\"uller-Kirsten\raisebox{0.6ex}{\small **}, 
D. K. Park\raisebox{0.6ex}{\small ***} and 
F. Zimmerschied\raisebox{0.6ex}{\small ****}}
\vspace{5mm}
\centerline{\it Department of Physics, University of Kaiserslautern,}

\centerline{\it 67653 Kaiserslautern, Germany}
\vspace{8mm}
\centerline {\large {\bf Abstract}}
\vspace{3mm}
\noindent
Some models allowing explicit calculation of periodic instantons and evaluation of
their action are studied with regard to transitions from classical to quantum 
behaviour as
the temperature is lowered and tunneling sets in.  It is shown that (beyond a
critical value of a coupling) the spin system considered acquires
a first order transition as a result of the field dependence of its effective 
mass,
whereas models with constant mass exhibit only second order transitions.

\vspace{5mm}

\vspace{0.5in}

Recently considerable progress was made in understanding macroscopic spin systems with
degenerate perturbation theory vacua.  This progress was achieved largely independently of
developments in field theory where very similar methods were motivated by the
realisation that topologically nontrivial field configurations play an important role
in our understanding of fundamental particle phenomena.  Thus instanton methods
well--known in theoretical particle physics for more than two decades appeared
in spin tunneling investigations about ten years later\cite{1}.  In either direction it took again
several years for topologically unstable classical field configurations,
which represent saddle points in path integrals, to be accepted as a convenient means
to understand decay processes, although in essence the configuration now known
as a bounce was already familiar to Langer\cite{2} several decades ago.  However,
there are very few theories which permit an explicit
calculation and investigation of such classical field configurations.  The best known example
to provide an instanton is quantum mechanics of the double--well
and sine--Gordon potentials.  This instanton is the vacuum instanton analogous to a
classical particle travelling in imaginary time with zero energy;  it is this instanton
which is responsible for the splitting of the degenerate perturbation theory
vacua into the two lowest quantum states.  Tunneling at higher perturbation theory
states is mediated by periodic instantons whose classical analogs are instantons
travelling with nonzero energy.  Peculiarly the study of these and their
stability in double--well and sine--Gordon theories\cite{3,4} began only about ten years ago,
and were called thermons\cite{5} in independent spin tunneling contexts.  Since then
progress in the study of spin systems uncovered a host of other model
theories permitting explicit evaluation of periodic instantons and their
investigation.

Very recently spin systems
aroused yet again new interest with the
discovery
\cite{6} that they provide examples which
exhibit first--order
phase transitions of which simple examples were not known previously.
In view of the possibility of experimental verification of such a
transition in decay rates of certain spin systems, and their interpretation
as a crossover from classical to quantum behaviour, such systems are also
of fundamental interest.  Very few models are known which
allow an explicit and analytic investigation of phase transitions so that
these few models serve as very instructive prototypes and are
of interest beyond their immediate area of relevance.

The characteristic way in which phase transitions appear in tunneling processes has been
worked out by Chudnovsky\cite{5}.  In fact the sharp first--order transition
is there shown to appear in the plot of action versus temperature
and is completely analogous to the plot of free enthalpy versus pressure of a
van der Waals gas whose equation of state plotted as pressure versus volume corresponds
to the plot of period (of the periodic instanton) versus energy in the consideration
of spin systems.

In the following we consider some models permitting explicit calculation of
periodic instantons and the corresponding evaluation of the action.
We first consider the well--known double--well\cite{7} and sine--Gordon\cite{8} theories
and demonstrate that these theories exhibit only second--order transitions. We then
consider a spin model with $XOY$ easy--plane anisotropy without an applied
magnetic field\cite{9} and demonstrate that this is a model exhibiting a first--order
transition.  We can clearly pin--point the reason for the appearance of this
first--order transition here as compared to the previous models  and attribute
it to the field dependence of the effective mass of the system which causes
the period of the periodic instanton to increase again after a certain critical value
of a coupling 
with increasing energy.  This is a very important point
which one can expect to appear in numerous other models,
also in the context of high energy physics, e.g. in models such as the Skyrme
model, which also possess field dependent masses.  

A uniaxial spin model
with an applied magnetic field has been considered in the work of
Chudnovsky and Garanin\cite{6} who also demonstrated that the crossover from classical
to quantum behaviour is, in fact, controlled by the magnetic field in their
model which has the effect of producing a shallow potential well
which effectively permits the period of the periodic instanton to increase after
a certain critical value.  In a subsequent   
work\cite{magphase} the model considered here --
which is different from the model of Chudnovsky and Garanin in the 
fieldless\cite{6} case
-- is investigated
with an additional applied magnetic
field, and a similar conclusion is arrived at.

As stated we begin with the discussion of periodic instantons of double--well and
sine--Gordon potentials and demonstrate that these lead only to smooth, i.e.
second order transitions.  In either case we use the notation and results of refs.\cite{7,8}.
We write the double--well potential
\begin{equation}
V(\phi ) = {\frac{\eta^2}{2}}{\left[\phi^2 - \frac{m^2}{\eta^2}\right]}^2
\label{1}
\end{equation}
Solving the Euclidean time classical equation in the usual
way one obtains the Jacobian elliptic function
$sn\left[b(k)\tau\right]) $
as the periodic instanton solution with period
\begin{equation}
P(E) = \frac{4}{b(k)} {\cal K}(k)
\label{2}
\end{equation}
where
\begin{equation}
b(k) = m\sqrt{\frac{2}{1+k^2}},\;\; k^2 = \frac{1-u}{1+u},\;\; u = \frac{\eta}{m^2}\sqrt{2E}
\label{3}
\end{equation}
Here $k$ is the elliptic modulus of the Jacobian elliptic functions, ${\cal K}(k)$ is their
quarter period, $m$ and $\eta $ are parameters of the potential and $E$ is the integration
constant which can be interpreted as the energy of the periodic instanton.  As is well--known,
in statistical mechanics the period is related to temperature $T$ through the relation
$P(E) = \frac{\hbar}{k_BT}$ where $k_B$ is Boltzmann's constant.  One can show that the
equation
\begin{equation}
\frac{dP(E)}{dE} = 0
\label{4}
\end{equation}
does not have a solution in the domain $0 < E < \frac{m^4}{2\eta^2} = E_0$ where $E_0$
is the energy of the pseudoparticle at the top of the barrier, also known as the sphaleron.
The monotonically decreasing behaviour of the period in this domain is shown in Fig.1(a).  
The thermon defined in ref.\cite{5} is a configuration travelling through one complete
period, whereas the periodic instanton, in keeping with its name, is defined
as a configuration over half the complete period.  Thus the action of the former
is twice that of the latter, and hence (with $\hbar = 1 = k_B$) is as shown in
ref.\cite{7}
\begin{equation}
S_T = \frac{E}{T} + \frac{8m^3}{3\eta^2}\sqrt{1+u}\left(E(k) - u{\cal K}(k)\right)
\label{5}
\end{equation}
where $E(k)$ is the complete elliptic integral of the second kind\cite{Shepard}.
The thermodynamic
action, i.e. that of the sphaleron at the top of the barrier, is correspondingly given by
\begin{equation}
S_0 = \frac{m^4}{2\eta^2}.\frac{1}{T}
\label{6}
\end{equation}
Fig.1(b) displays the behaviour of $S_T, S_0$ versus temperature $T$
for $m =1, \eta =1$. One can clearly see the typically smooth behaviour of
a second order transition from the thermal to the quantum regime as the
temperature is lowered. Conversely, we can argue that as the temperature is 
increased, the number of periodic instantons and antiinstantons or thermons
increases and the dilute gas approximation breaks down. The temperature 
corresponding to that of harmonic oscillations \cite{5} around the sphaleron
configuration is the critical temperature at which the periodic instantons and
antiinstantons condense and disorder the system.

In the case of the sine--Gordon potential written
\begin{equation}
V(\phi )= \frac{1}{g^2}\left[1 + \cos (g\phi)\right]
\label{7}
\end{equation}
one finds $arc\sin[ksn(\tau)]$ as the periodic instanton (cf. ref.\cite{8}) with period
\begin{equation}
P(E) = 4{\cal K}(k),\;\; k = \sqrt{1-\frac{g^2E}{2}}
\label{8}
\end{equation}
and again eq.(\ref{4}) 
can be shown not to possess a solution in the domain $0 < E < \frac{2}{g^2}$.
The thermon and thermodynamic actions are obtained as
\begin{equation}
S_T =\frac{E}{T} + \frac{16}{g^2}\left[E(k) - {k^{\prime}}^2{\cal K}(k)\right],\;\;S_0 =\frac{2}{g^2}.
\frac{1}{T}
\label{9}
\end{equation}
where $k^{\prime} = \sqrt{1-k^2} $ and $T$ is again the temperature.  
Again one finds a second order transition from the thermal to the quantum regime
very similar to that depicted in Fig. 1(b).

\begin{figure}
\vspace{-1cm}
\unitlength1cm
\begin{center}
\includegraphics[bb= 1.5cm 5.5cm 18cm 26.5cm,clip,scale=1]
{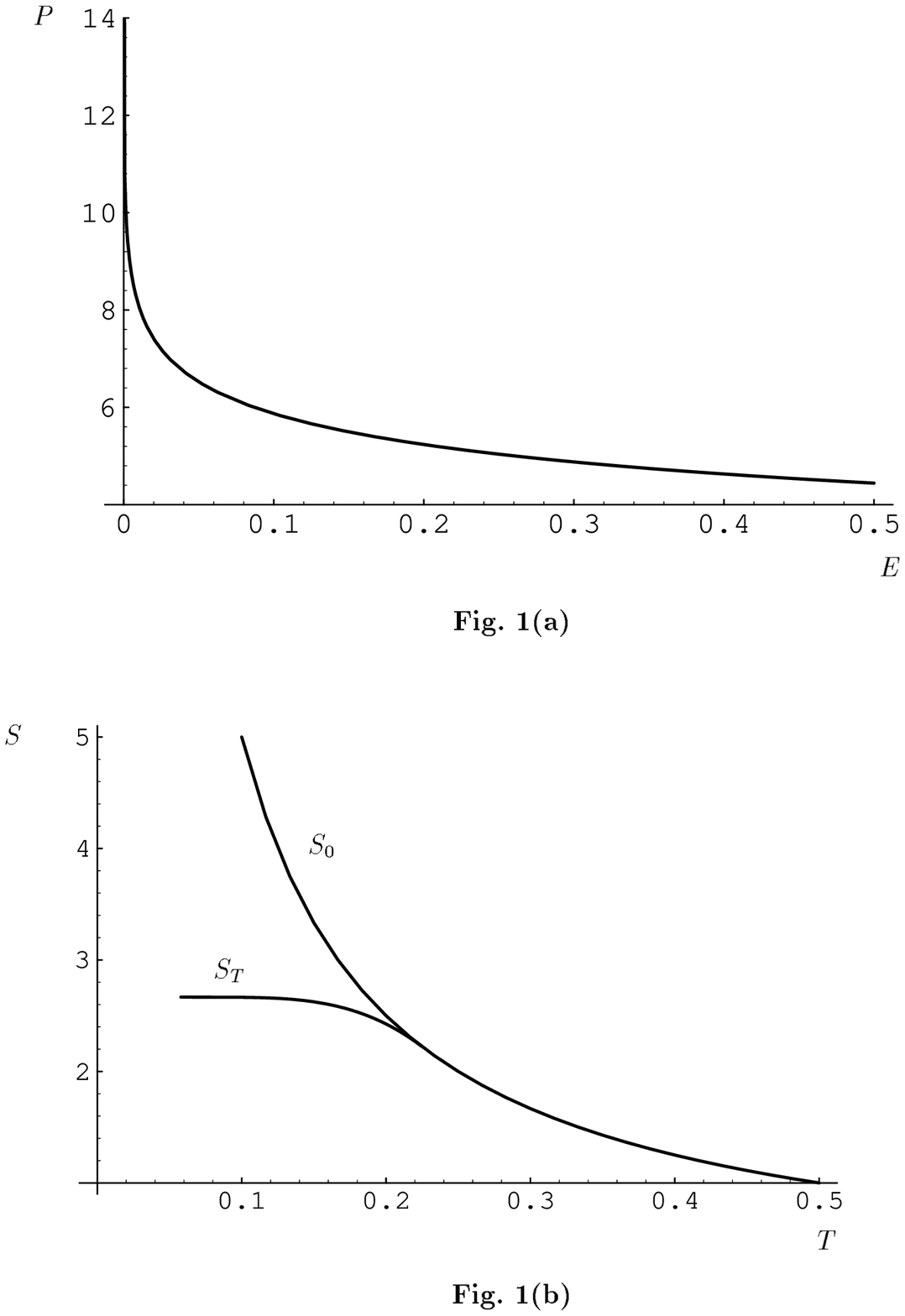}
\end{center}
\vspace{3mm}
\textbf{Fig.\ 1(a).} 
The period of the periodic instanton of the 
double--well potential as a function of energy $E$ for
$\eta = 1, m = 1$; \textbf{(b)} the thermodynamic and thermon
actions $S_0, S_T$ as functions of temperature $T$
demonstrating the smooth second--order transition
for the same values of parameters.
\end{figure}

We now consider the particular spin tunneling model without applied
magnetic field we investigated earlier in ref.\cite{9} and which allows
the explicit calculation of periodic instantons as well as the evaluation of 
their action, and we
show that this  model gives rise to a first order transition.
The model is described in refs.\cite{1,9,10,11,12} and a lot of further 
information can be
found in ref.\cite{13}. We therefore restrict ourselves here to the essential 
aspects
of relevance in our present consideration.  With the help of the coherent--state
path integral the theory defined by the original Hamiltonian
, i.e.\ $H=K_1S_z^2 + K_2 S_y^2$, can be shown to
lead to an equivalent effective continuum theory with Hamiltonian
\begin{equation}
H = \frac{p^2}{2m(\phi )} + V(\phi )
\label{10}
\end{equation}
and
\begin{equation}
m(\phi ) = \frac{1}{2K_1(1-\lambda \sin^2\phi)},\;\; V(\phi ) = K_2s(s+1)\sin^2\phi
\label{11}
\end{equation}
where $\lambda \equiv \frac{K_2}{K_1}$ is assumed to be less than $1$,
and the spin eigenvalue $s$ is assumed to be large (i.e. much larger than 1). We see that
the potential is again a periodic potential as in the sine--Gordon theory, but now
the effective mass is field dependent. It will be seen that this field dependence is
crucial in leading to a first order transition from the classical to the quantum regime.
In ref.\cite{9} it is shown that the classical equation associated with the model
possesses the following periodic instanton configuration
\begin{equation}
\phi = \arcsin{\left[\frac{1-k^2 sn^2(\omega\tau|k)}{1-\lambda k^2 sn^2(\omega\tau|k)}
\right]}^{\frac{1}{2}}
\label{12}
\end{equation}
where $sn(\omega\tau|k) $ again denotes the Jacobian elliptic function of modulus $k$,
and
\begin{equation}
k=\sqrt{\frac{n^2-1}{n^2-\lambda}},\;\;n^2 = \frac{K_2s(s+1)}{E},\;\; 
\omega={\omega}_0\sqrt{1-\frac{\lambda}{n^2}},\;\;{\omega}_0^2=4K_1K_2s(s+1)
\label{13}
\end{equation}  
Again we consider first the energy dependence of the period 
$P(E)$ which is given by
\begin{equation}
P(E) = \frac{2}{\sqrt{K_1}}\frac{1}{\sqrt{K_2s^2 - E\lambda}}{\cal K}(k),\;\;
k\equiv \sqrt{\frac{K_2s^2-E}{K_2s^2-E\lambda}}
\label{14}
\end{equation}
We consider again eq.(\ref{4}) and enquire about a nontrivial solution corresponding to an
energy $E$ in the domain $0 <  E < E_{0}\equiv K_2s^2$. Such a solution would
violate the monotonically decreasing behaviour of $P(E)$ observed in the
earlier examples. Using the formula
\begin{equation}
\frac{d{\cal K}(k)}{dk} = \frac{1}{k}\left(\frac{E(k)}{{k^{\prime}}^2} - {\cal K}(k)\right)
\label{14b}
\end{equation}
one obtains for eq.(\ref{4})
\begin{equation}
{\cal K}(k) - \frac{K_2s^2}{E}E(k) = 0
\label{15}
\end{equation}
The solution of this equation can be investigated both numerically and with
approximation analytically. In the numerical procedure we choose $K_1 = 1$
(thus $\lambda = K_2$), and $s = \sqrt{1000}$, and calculate the energy $ E_1$ of the minimum of
$P(E)$ for $K_2$ varying from $0.9$ downwards.  The results in Table 1
show that $E_1$ reaches the maximum energy $E_0$
when $K_2$ approaches $0.5$.

\vspace{5mm}
\centerline{\bf Table 1}

\centerline{Numerical determination of critical value $E_1$}
\vspace{2mm}
\begin{tabular}{|c|c|c|}
\hline
$K_2$&Energy $E_1$ at minimum of $P(E)$ &$E_0 = K_2s^2$ (Barrier height)\\
\hline
$0.9$&$336.81$&$900$\\ \hline
$0.7$&$430.75$&$700$\\ \hline
$0.6$&$446.36$&$600$\\ \hline
$0.51$&$496.67$&$510$\\ \hline
$0.501$&$500$&$501$\\ \hline
\end{tabular}
\vspace{0.4in}

\noindent
Thus the critical value of $K_2$ is $\frac{1}{2}$ implying that smaller values of $K_2$
correspond to unphysical values of $E$.  We obtain the same condition
analytically using the expansions of ${\cal K}(k)$ and $E(k)$ in rising powers
of $k^2$ around $\frac{\pi}{2}$. 
Taking into account terms of $O(k^2)$ in these expansions, one finds two
possible solutions, i.e.
\begin{equation}
E = K_2s^2\;\;\; or\;\;\; E = \frac{3K_2s^2}{1 + 4\lambda}
\label{16}
\end{equation}
Since $E < E_0$, the nontrivial solution is obtained for
$$
\frac{3K_2s^2}{1+4\lambda} < K_2s^2, \;\; i.e.\;\; \lambda > \frac{1}{2}
$$
in agreement with the numerical finding.  

\begin{figure}
\vspace{-1cm}
\unitlength1cm
\begin{center}
\includegraphics[bb= 1.5cm 5.5cm 18cm 26.5cm,clip,scale=1]
{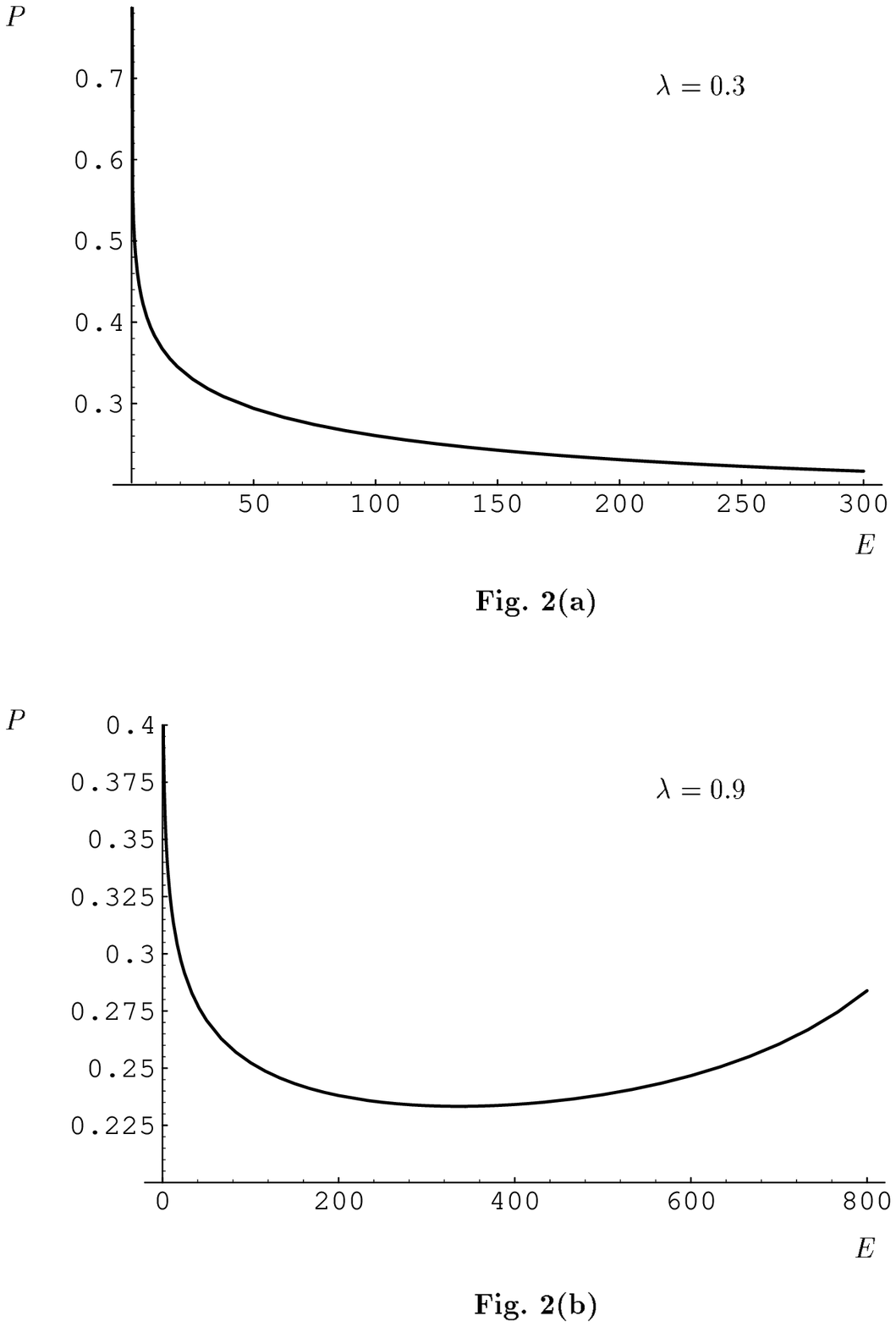}
\end{center}
\vspace{5mm}
\textbf{Fig.\ 2.} The period of the periodic instanton in the fieldless spin 
model as a function of energy $E$ for $s = \sqrt{1000}$ and $K_1 = 1$: 
In \textbf{(a)} for $\lambda = 0.3$ 
and in \textbf{(b)} for $\lambda = 0.9.$
\end{figure}

\begin{figure}
\unitlength1cm
\begin{center}
\includegraphics[bb= 1.9cm 10.5cm 20cm 20.5cm,clip,scale=1]
{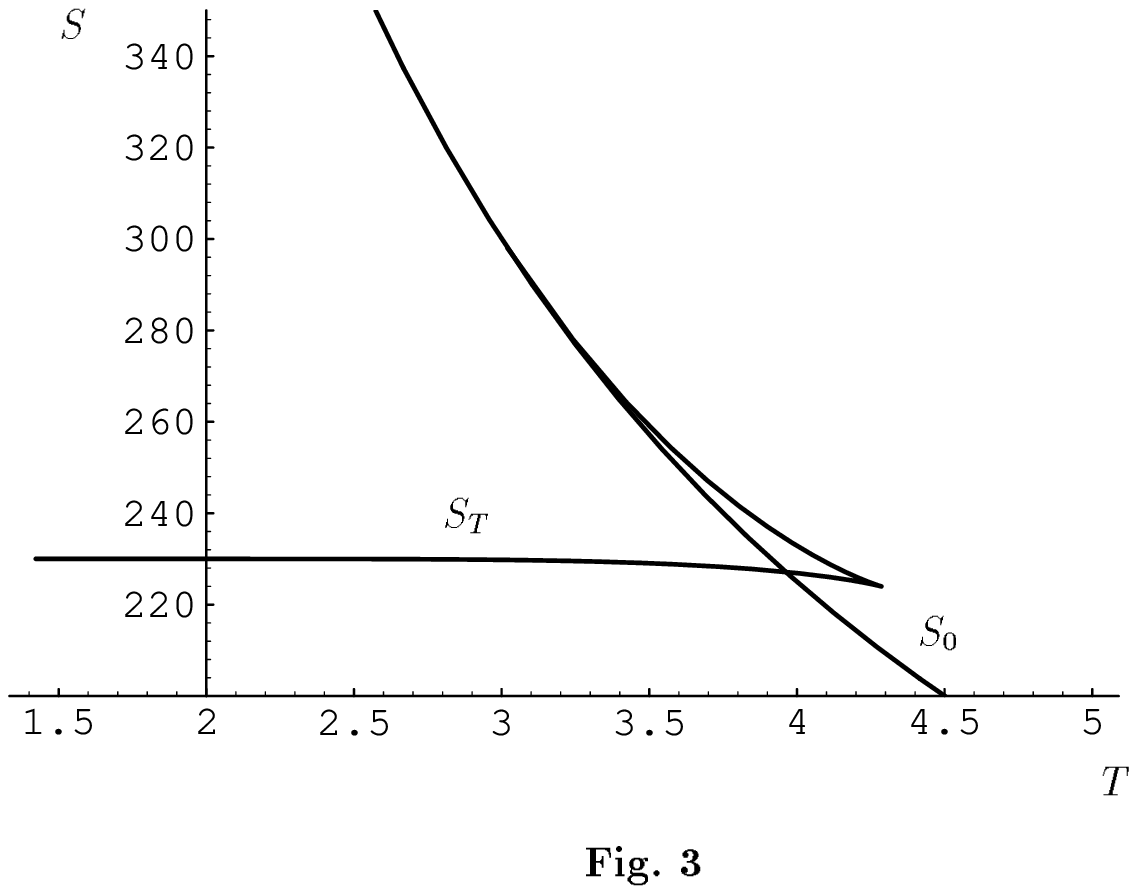}
\end{center}
\vspace{3mm}
\textbf{Fig.\ 3.} 
The thermodynamic and thermon actions $S_0, S_T$ of the fieldless spin model as 
functions of temperature $T$ for $\lambda = 0.9$ demonstrating the
first order transition.
\end{figure}

Continuing in the present case as in the earlier examples we have the
thermodynamic action
\begin{equation}
S_0 = \frac{K_2s^2}{T}
\label{17}
\end{equation}
and the thermon action (which is twice the action of the periodic instanton
given in ref.\cite{9})
\begin{equation}
S_T=\frac{E}{T} + 2W, \;\; W = \frac{\omega}{\lambda K_1}\left[{\cal K}(k) -
(1-k^2\lambda)\Pi(k^2\lambda, k)\right]
\label{18}
\end{equation}
where $\omega = 2\sqrt{K_1(K_2s^2 - E\lambda)}$ and $\Pi(k^2\lambda,k)$ is
the complete elliptic integral of the third kind.
Fig. 2(a) shows the monotonically decreasing behaviour of the period $P(E)$
as a function of energy for $s =\sqrt{1000}, K_1 = 1$ and $\lambda = 0.3$,
i.e. less than the critical value $\frac{1}{2}$.
In Fig. 2(b) we plot the period for the corresponding case with 
$\lambda = 0.9$, i.e. larger than $\frac{1}{2}$. The rising behaviour
of the period after reaching the critical energy value of $336.81$
results from the increasing importance 
of the field dependence of the effective mass. It is this kind of 
rising behaviour of the period which is necessary in order to
generate a first order transition.  We see this in the corresponding
plots of the thermodynamic and thermon actions shown in Fig.3,
again plotted for the value $0.9$ of $\lambda$, in which only the two lowest 
branches, marked $S_0$ and $S_T$, are physical.
 
In the above we have therefore made a very general observation.
The regulation of the transition from the classical to the quantum
regime by variation of the applied magnetic field as observed
by Chudnovsky and Garanin\cite{6} can also be achieved through
variation of the field dependence of the effective mass.  We expect
this characteristic to show up also in field theory models such
as the Skyrme model and its variants.
\vspace{8mm}

\noindent
{\bf Acknowledgement}
\noindent
D.-K. Park  acknowledges support of the Deutsche Forschungsgemeinschaft (DFG).

\vspace{5mm}
\raisebox{0.6ex}{\small *} Permanent address: 
Department of Physics, Shanxi University, Taiyuan 030006, P.R. China.
Electronic address: liangjq@sun.ihep.ac.cn

\raisebox{0.6ex}{\small **} Electronic address: mueller1@physik.uni-kl.de

\raisebox{0.6ex}{\small ***} Permanent address: Department of Physics, 
Kyung Nam University, Masan 631--701, Korea.
Electronic address: dkpark@chep5.kaist.ac.kr

\raisebox{0.6ex}{\small ****} Electronic address: zimmers@physik.uni-kl.de

\end{document}